\documentstyle[12pt]{article}
\newcommand{\mysection}[1]{\section{#1}\setcounter{equation}{0}}

\def\bea{\begin{eqnarray}} 
\def\eea{\end{eqnarray}}
\def\beann{\begin{eqnarray*}} 
\def\eeann{\end{eqnarray*}}
\def\beq{\begin{equation}} 
\def\eeq{\end{equation}}
\def\ba{\begin{array}} 
\def\ea{\end{array}}

\def\5{\bar }  
\def\6{\partial } 
\def\7{\hat }

\def\agh{\mbox{agh}\,}

\def\suba{{\underline a}} \def\subb{{\underline b}} \def\subc{{\underline c}}
\def\resta{{\hat a}}  \def\restc{{\hat c}}

\def\ep{\epsilon}
\def\UD#1#2{\stackrel{(#1)}{#2}}
\def\da{{\dot\alpha}}
\def\dsl{\not\!\partial }
\def\Ii{{\mbox{i}}}

\begin{document}

\begin{flushright}
ITP--UH--07/98\\
hep-th/9804153\\
\end{flushright}

\begin{center}
{\LARGE Deformations of global symmetries in the extended
antifield formalism}
\end{center}

\begin{center}
{\large
Friedemann Brandt}
\end{center}

\begin{center}{\sl
 Institut f\"ur Theoretische Physik,
 Universit\"at Hannover,\\
 Appelstra\ss e 2,
 D--30167 Hannover, Germany\\
 E-mail: brandt@itp.uni-hannover.de
}\end{center}

\begin{abstract}
It is outlined how deformations of field theoretical rigid symmetries can 
be constructed and classified by cohomological means in the extended 
antifield formalism. Special attention is devoted to deformations referring
only to a subset of the rigid symmetries of a given model and leading
to a nontrivial extension of the graded Lie algebra associated with that 
subset. The method is illustrated for a D=4, N=2 supersymmetric model where 
the central extension of the supersymmetry algebra emerges via a deformation. 
Deformations of gauge fixed actions with a BRST symmetry are discussed too 
and illustrated by the Curci-Ferrari model.
\end{abstract}

\mysection{Introduction}

A problem often met in field theory is to what degree a given 
action functional can be nontrivially deformed while keeping 
some of its symmetries.
A particularly interesting issue is whether
the symmetry transformations themselves can be deformed
in a nontrivial way, i.e.\ whether there are simultaneous
deformations of the action and its symmetries.

Deformations of this sort can be studied systematically by
cohomological methods in the spirit
of Gerstenhaber's approach to deformation theory \cite{gerstenhaber}.
This was first
described in \cite{bh} (see also \cite{henmoscow,stasheff})
for gauge symmetries in the framework
of the standard antifield formalism \cite{bv,henteit,report}.
The inclusion of
rigid (= global) symmetries was roughly sketched more recently 
in \cite{bhw} within an extended antifield formalism.
The aim of this work is to develop the latter approach more thoroughly,
with special attention to deformations which are
required to maintain only (a deformed version of) a 
{\em subset} of the rigid symmetries of a given model.

The restriction to a subset of
the rigid symmetries is a typical situation, as
often it is neither possible nor desirable to
keep all the rigid symmetries when deforming a field theory
because that may constrain the sought deformations too much.
We shall thus base the deformation theory on
an extended antifield formalism which involves only a ``closed''
subset of rigid symmetries. The ``closure'' of the subset requires
that the graded commutator algebra of the rigid symmetries under study
closes in the soft (field theoretical) sense,
i.e.\ up to gauge transformations and on-shell trivial
symmetries\footnote{In order to set up the extended antifield
formalism, it may be necessary to include also
``symmetries of higher order'' \cite{bhw,bhw1}.}.
In other words, a closed subset of rigid symmetries
forms a subalgebra (in the soft sense)
of the graded commutator algebra of all the rigid symmetries.

When one applies the extended antifield formalism to
study deformations of such a subset of rigid symmetries,
one may encounter a ``subtlety''. Namely, a deformation may turn a
subset of rigid symmetries which is closed in the soft sense 
into an open one.
That is, it can happen that the deformed
commutator algebra involves symmetries which did not
occur in the undeformed one. These additional 
symmetries are not ``new'' ones which are
introduced through the deformation. Rather, they are
present already in the original (undeformed) model. 
The subtlety is that usually it is
not clear from the outset which additional symmetries of the 
original model can show up in the deformed commutator algebra. 
In particular, this may depend on the deformation itself.

Hence, the property of
a subset of rigid symmetries to be a closed one
is not necessarily preserved by
deformations. This is actually an interesting phenomenon as it is
related to extensions of the
(graded) Lie algebra associated with the
commutator algebra of the subset of rigid symmetries under study.
Important examples are central extensions of
extended supersymmetry algebras \cite{hls}.
As an illustration, we shall discuss a simple four dimensional
N=2 supersymmetric model for a hypermultiplet \cite{Fa,So}
where the central extension of the
supersymmetry algebra arises indeed via a deformation of the model.

The antifield formalism serves in this context as a tool 
that allows one to formulate the deformation theory conveniently
in cohomological terms. Ghost fields are not
dynamical in this approach (in particular,
they are not paired with antighost fields), in contrast
to their counterparts in the quantum field theoretical context.
Nevertheless, the formalism applies also to gauge fixed action
functionals which contain dynamical ghost and antighost fields. 
This application just requires a slight change of the point of view
as compared to the one familiar from quantum field theory. 
Namely, the gauge fixed action simply takes
the role of a classical action. Accordingly, the dynamical ghost 
and antighost fields occurring in the gauge fixed action are 
counted among the classical fields,
and the BRST symmetry of the gauge fixed action 
counts among the rigid symmetries. In particular
this allows one to investigate deformations of the
BRST symmetry after fixing the gauge.
We shall discuss and illustrate this particular application 
in some detail in the Curci-Ferrari model \cite{CF,O,kostas}.

The paper has been organized as follows.
Section \ref{symm} summarizes basic properties of
global and local symmetries in Lagrangian field theory which
are used later on. Then
the extended antifield formalism and the construction
and properties of the extended BRST differential are
briefly reviewed in sections 
\ref{AF} and \ref{KT}. The systematic approach
to the deformation problem is described in section \ref{DEF}.
Sections \ref{HM} and \ref{CF} contain the
examples mentioned above, i.e.\ the
hypermultiplet of N=2 supersymmetry and the Curci-Ferrari model.
The paper is ended with some concluding remarks
in section \ref{CON}.

\mysection{Global and local symmetries}\label{symm}

We shall first briefly summarize the definition and some 
properties of continuous rigid and gauge symmetries in
Lagrangian field theories, following the presumably most
popular approach based on the action (alternatively one
can define rigid and gauge symmetries
on the level of the field equations,
via conserved currents and Noether identities respectively).
We shall thus consider Lagrangian field theories
which derive from an action functional for a set of fields $\phi^i(x)$,
\beq 
S_{\mbox{class}}[\phi]=\int d^nx\, L(x,[\phi]),
\label{Sact}\eeq
where $L(x,[\phi])$ is a Lagrangian
constructed of the fields and their partial derivatives%
\footnote{Here and in the following, $[\phi]$ denotes collectively
dependence on the fields and on their derivatives. 
In more precise mathematical terms, $\phi^i$, $\6_\mu\phi^i$, 
$\6_\mu\6_\nu\phi^i$, \dots
are to be understood as local coordinates
of a jet space, and $\phi^i(x)$
as sections of the jet bundle over an $n$-dimensional 
base manifold (``spacetime'') with local coordinates 
$x^\mu$ ($\mu=1,\dots,n$).
The arguments of $L(x,[\phi])$ indicate that the Lagrangian
may (but, of course, need not) depend explicitly on the $x^\mu$.}.
The field equations
(equations of motion) derive via the variational principle
from $S_{\mbox{class}}$, i.e.\ they are the corresponding Euler-Lagrange
equations.

A continuous rigid symmetry of an action (\ref{Sact}) is generated by
transformations of the fields with a constant
infinitesimal parameter $\varepsilon$,
\beq 
\phi^i\rightarrow \tilde\phi^i=\phi^i
+\varepsilon\, G^i(x,[\phi]),
\quad \varepsilon=\mbox{constant},
\label{S2a}\eeq
such that $L(x,[\tilde\phi])$ differs from $L(x,[\phi])$
to first order in $\varepsilon$
at most by a total derivative,
\beq
L(x,[\tilde\phi])
=L(x,[\phi])
+\varepsilon\,\6_\mu k^\mu(x,[\phi])
+O(\varepsilon^2).
\label{S3}\eeq

A gauge symmetry of an action (\ref{Sact}) is defined
similarly, with the important difference
that it involves, instead of a constant parameter, 
an additional field $\lambda=\lambda(x)$ (i.e.\ a field
which does not occur in the Lagrangian). It is
generated by infinitesimal transformations of the form
\beq 
\phi^i\rightarrow \tilde\phi^i=\phi^i
+ \sum_{k\geq 0}r^{i\,\mu_1\dots\mu_k}(x,[\phi])\,
\6_{\mu_1}\dots\6_{\mu_k}\lambda
\label{S4}\eeq
such that $L(x,[\tilde\phi])$ and $L(x,[\phi])$ differ
to first order in $\lambda$
at most by a total derivative, 
\beq
L(x,[\tilde\phi])=L(x,[\phi])
+\6_\mu h^\mu(x,[\phi,\lambda])+O(\lambda^2).
\label{S5}\eeq
This invariance condition must hold for an
unconstrained field $\lambda$, i.e., it must neither impose
a differential equation for $\lambda$, nor determine $\lambda$
in terms of the fields $\phi^i$ and their derivatives
(otherwise $\lambda$ would turn into a function 
of the $x^\mu$, $\phi^i$ and their derivatives and thus
(\ref{S4}) would reduce to a rigid symmetry of the form (\ref{S2a})).

Now, the above standard definitions do not yet characterize
symmetries satisfactorily for our purpose. An important ingredient,
underplayed in many textbooks, is still missing:
the distinction between trivial and nontrivial symmetries. For instance,
consider transformations (\ref{S2a}) and (\ref{S4}) with
\bea
& & G^i(x,[\phi])=E^{ij}(x,[\phi])\,\frac{\hat\6L}{\hat\6\phi^j}
\label{S10}\\
& &\sum_{k\geq 0}r^{i\,\mu_1\dots\mu_k}(x,[\phi])\,
\6_{\mu_1}\dots\6_{\mu_k}\lambda =
E^{ij}(x,[\phi,\lambda])\,\frac{\hat\6L}{\hat\6\phi^j}
\label{S11}\eea
where $\hat\6L/\hat\6\phi^i$ is the Euler-Lagrange derivative
of the Lagrangian with respect to $\phi^i$,
\beq
\frac{\hat\6L}{\hat\6\phi^i}=\frac{\6L}{\6\phi^i}
-\6_\mu\frac{\6L}{\6(\6_\mu\phi^i)}+\dots\ ,
\label{S7}\eeq
and $E^{ij}$ are any functions which are only required to be
graded antisymmetric
in their indices,
\beq
E^{ij}=-(-)^{\epsilon_i\,\epsilon_j}E^{ji},
\label{S12}\eeq
where $\epsilon_i$ is the Grassmann parity of $\phi^i$.
It is easily verified that (\ref{S10}) and (\ref{S11})
give rigid and gauge symmetries, satisfying (\ref{S3}) and (\ref{S5})
respectively, for any choice of
$E^{ij}$ fulfilling (\ref{S12}).
Such symmetries are examples of trivial symmetries
which may be called ``on-shell trivial symmetries''
(the terminology reflects that the symmetry transformations
vanish for every solution of the field equations, as the latter
read $\hat\6L/\hat\6\phi^i=0$).
More general trivial symmetries of this type are obtained from 
(\ref{S10}) and (\ref{S11}) 
when $E^{ij}$ are differential operators of the form
$E^{ij}=\sum e^{ij\,\mu_1\dots\mu_k}
\6_{\mu_1}\dots\6_{\mu_k}$ with properties
generalizing (\ref{S12}) appropriately.

In addition to on-shell trivial symmetries, 
there is a second type of trivial rigid symmetries whenever
the action possesses a true
gauge symmetry, i.e.\ a gauge symmetry which is
not on-shell trivial. Indeed, in that case the action 
has automatically infinitely many further rigid symmetries which
are to be considered as trivial too, even though
they are not on-shell trivial.
These additional trivial rigid symmetries arise from
nontrivial gauge transformations (\ref{S4})
by replacing there $\lambda$ with $\varepsilon\, f(x,[\phi])$,
where $f(x,[\phi])$ is any function of the fields and their
derivatives. Indeed, as (\ref{S5}) holds for any $\lambda$,
such a replacement results in a transformation (\ref{S2a}) 
satisfying (\ref{S3}) with 
\beq G^i(x,[\phi])=\sum_{k\geq 0}r^{i\,\mu_1\dots\mu_k}(x,[\phi])\,
\6_{\mu_1}\dots\6_{\mu_k}f(x,[\phi]).
\label{S14}\eeq

Hence, every action has infinitely many trivial gauge and rigid 
symmetries.
Gauge and rigid symmetries are therefore best defined as
{\em equivalence classes} where two symmetries 
are called equivalent when they differ by a trivial
symmetry (or by irrelevant redefinitions of the respective
$\varepsilon$ and $\lambda$, i.e.\ by 
multiplications of $\varepsilon$
and $\lambda$ with arbitrary constants and field dependent
functions respectively\footnote{Clearly, two symmetries
differing only through such redefinitions 
are to be identitied, as (\ref{S3}) and (\ref{S5}) must
hold for arbitrary constant parameters $\varepsilon$
and unconstrained fields $\lambda$.}).
One can then introduce the concept of a basis of
symmetries, containing  
one representative of each nontrivial equivalence class.
We shall characterize such bases for the gauge and rigid symmetries 
through operations
$\{\delta_\alpha\}$ and $\{\Delta_a\}$ respectively, which are
related to symmetry
transformations (\ref{S2a}) and (\ref{S4}) according to
\bea
\delta_\alpha\phi^i&=&R^i_\alpha(x,[\phi])\equiv
\sum_{k\geq 0}r_\alpha^{i\,\mu_1\dots\mu_k}(x,[\phi])\,
\6_{\mu_1}\dots\6_{\mu_k}
\label{S15}\\
\Delta_a\phi^i&=&G^i_a(x,[\phi]).
\label{S16}\eea

As the graded commutator of two infinitesimal symmetry transformations
is automatically again an infinitesimal symmetry transformation (due
to the derivation property of
infinitesimal transformations),
there is always a graded commutator algebra associated with such bases.
However, due to the presence of trivial symmetries, this graded
commutator algebra is in general a quotient algebra because, in general,
the graded commutator of two elements of the basis can be expressed
in terms of elements of the same basis only up to trivial symmetries.
In particular, the general form of the graded commutator
of any two elements of a 
basis of infinitesimal rigid symmetry transformations is thus
\beq
\mbox{{\bf [}}\, \Delta_a\, ,\, \Delta_b\mbox{{\bf ]}}\,\phi^i
= f_{ab}^c \Delta_c\phi^i
+R^i_\alpha f^\alpha_{ab}(x,[\phi])
+E^{ij}_{ab}(x,[\phi])\,\frac{\hat\6L}{\hat\6\phi^j}
\label{S17}\eeq
where the graded commutator of two objects $A$ and $B$ is defined by
means of their Grassmann parities $\epsilon(A)$ and $\epsilon(B)$ through
\beq
\mbox{{\bf [}}\, A\, ,\, B\mbox{{\bf ]}}=
AB-(-)^{\epsilon(A)\epsilon(B)}BA.
\eeq
In (\ref{S17}), $f_{ab}^c$ are constant coefficients which
are the structure constants of a graded
Lie algebra (as a consequence of
$\mbox{{\bf [}}\,\mbox{{\bf [}}\Delta_a,\Delta_b\mbox{{\bf ]}},
\Delta_c\mbox{{\bf ]}}+cyclic=0$), while
$f^\alpha_{ab}(x,[\phi])$ and $E^{ij}_{ab}(x,[\phi])$
are in general field dependent functions and operators
appearing in trivial rigid symmetries as described above,
cf.\ (\ref{S14}) and (\ref{S10}).

\mysection{Extended antifield formalism}\label{AF}

We shall now recall the basic features of the extended antifield
formalism and fix our notation and conventions. For
simplicity, we shall concentrate on
the case that the gauge transformations (if any) are irreducible and 
that only ordinary rigid symmetries are
present or needed, but no rigid symmetries of higher order in the
terminology of \cite{bhw}. The general case is a straightforward
extension of this one. As mentioned already, the extended
antifield formalism can be established for any closed
subset of rigid symmetries \cite{bhw}. 
When higher order rigid symmetries are absent,
a closed subset is simply
a subset $\{\Delta_\suba\}$ of a basis $\{\Delta_a\}$ of nontrivial 
rigid symmetries such that, in the notation of the previous section,
\beq 
\{\Delta_a\}=\{\Delta_\suba\ ,\, \Delta_\resta \}
\quad ,\quad 
f_{\suba\subb}^\restc=0\quad \forall\, \suba,\subb,\restc \ .
\label{subset}\eeq
$f_{\suba\subb}^\restc=0$ requires that the graded commutator algebra
of the $\Delta_\suba$ is a subalgebra of (\ref{S17}) in the ``soft''
sense, i.e.\ with respect to the quotient structure ``modulo trivial 
rigid symmetries''.

The fields and antifields of the standard antifield formalism 
are denoted by $\Phi^A$ and $\Phi^*_A$ where $\{\Phi^A\}$
contains the ``classical'' fields $\phi^i$, i.e.\ the fields occurring 
in the ``classical'' action (\ref{Sact}) under study, and the ghost fields
$C^\alpha$ corresponding to the nontrivial gauge symmetries of this action%
\footnote{As mentioned in the
introduction, the ``classical'' action may be actually a gauged fixed one.
Ghost and antighost fields occurring in such an action count
among the $\phi^i$ and must not be confused with the $C^\alpha$,
see section \ref{CF} for an example.}.
The extended antifield formalism, restricted to the
subset $\{\Delta_\suba\}$, contains in addition a constant
(``global'') ghost $\xi^\suba$ for each $\Delta_\suba$.
These global ghosts have ghost number
1 and Grassmann parity opposite to the corresponding rigid
symmetries. It is also very convenient (though not
necessary in principle) to accompany each $\xi^\suba$ with a constant
antifield $\xi^*_\suba$. The latter has ghost number $(-2)$ and
Grassmann parity opposite to $\xi^\suba$. In particular this allows
one to set up the extended antifield formalism
through an extended master equation of the form
\beq
(S,S)=0
\label{me}\eeq
where $(\ ,\ )$ is an extended antibracket defined by
\bea
(X,Y)&=&
 \frac{\partial^R X}{\partial\xi^\suba}\,
 \frac{\partial^L Y}{\partial\xi^*_\suba}-
 \frac{\partial^R X}{\partial\xi^*_\suba}\,
 \frac{\partial^L Y}{\partial\xi^\suba}
\nonumber\\
& & 
 +\int d^n x \left[
 \frac{\delta^R X}{\delta\Phi^A(x)}\,
 \frac{\delta^L Y}{\delta\Phi^*_A(x)}-
 \frac{\delta^R X}{\delta\Phi^*_A(x)}\,
 \frac{\delta^L Y}{\delta\Phi^A(x)}\right].
\label{ab}\eea
Here superscripts $R$ and $L$ indicate right and left derivatives
respectively.
The extended antibracket is defined in the
space of local functionals of the form
\beq
\Gamma[\Phi,\Phi^*,\xi]+M^\suba(\xi)\,\xi^*_\suba
\label{space}\eeq
where $\Gamma[\Phi,\Phi^*,\xi]$ is the spacetime integral
of a local function of the fields
and antifields which may depend on the global ghosts but not
on the global antifields, and $M^\suba(\xi)$ is a polynomial
in the global ghosts (note: $M^\suba(\xi)\,\xi^*_\suba$ does not
involve a spacetime integration).
The solution $S$ of the extended master equation is
a functional with ghost number 0 of the form (\ref{space}). It
contains the classical action,
and encodes its gauge symmetries and the subset 
$\{\Delta_\suba\}$ of its rigid symmetries,
as well as the graded commutator algebra of these
symmetries. In addition one often imposes
that $S$ be real. One then needs consistent conventions for
complex conjugation.
We denote complex conjugation by a bar, and use the convention
(familiar from supersymmetry, see, e.g., \cite{wb})
that complex conjugation of products involves a sign factor
depending on the Grassmann parities,
\beq 
\overline{(XY)}=(-)^{\ep_X\ep_Y}\bar X\, \bar Y.
\label{cc}\eeq
The complex conjugate of an antifield $\Phi^*$ equals minus the
antifield of the complex conjugate of $\Phi$ (independently of
the Grassmann parity of $\Phi$),
\beq
\overline{(\Phi^*)}=-(\bar \Phi)^*\quad
\forall\, \Phi\in\{\Phi^A,\xi^\suba\}.
\eeq
For instance, with these conventions, the antifield of a real field is
purely imaginary.

To describe and compute $S$, it is useful to expand it
in the antifield number ($\agh$).
The latter vanishes for the fields, and equals minus the ghost number
for the antifields,
\bea 
& \agh \phi^i=\agh C^\alpha=\agh\xi^\suba=0 &
\nonumber\\
& \agh \phi^*_i=1,\quad \agh C^*_\alpha=\agh \xi^*_\suba=2. &
\label{agh}\eea
The expansion of $S$ is denoted by
\beq
S=\sum_{k\geq 0}S_k\ ,\quad \agh S_k=k\ .
\eeq
Here $S_0$ is the classical action,
\beq
S_0=S_{\mbox{class}}[\phi]\ .
\label{S0}\eeq 
$S_1$ encodes both the gauge transformations
and the subset of the rigid symmetries under study,
\beq
S_1=-\int d^nx \left(R^i_\alpha C^\alpha
+\xi^\suba\Delta_\suba\phi^i\right)\phi^*_i
\label{S1}\eeq
where we used the notation of the previous section.
$S_2$ encodes the graded commutator algebra of the
gauge symmetries and the subset of rigid symmetries
under study, and thus in particular
the subalgebra of (\ref{S17}) referring to $\{\Delta_\suba\}$,
\beq
S_2= 
\frac 12\, \xi^\subb\xi^\suba\tilde f_{\suba\subb}^\subc\xi^*_\subc
+\int d^nx\, \xi^\subb\xi^\suba 
\left(\frac 12\,\tilde f^\alpha_{\suba\subb} C^*_\alpha
+\frac 14\, \phi^*_i\tilde E^{ij}_{\suba\subb}\phi^*_j+\dots\right),
\label{S2}\eeq
where $\tilde f_{\suba\subb}^\subc$, $\tilde f^\alpha_{\suba\subb}$,
$\tilde E^{ij}_{\suba\subb}$ coincide with
$f_{\suba\subb}^\subc$, $f^\alpha_{\suba\subb}$,
$E^{ij}_{\suba\subb}$ in (\ref{S17})
up to signs which follow from the formulae
(e.g., $\tilde f_{\suba\subb}^\subc
=(-)^{\ep_\subb+1} f_{\suba\subb}^\subc$
where $\ep_\subb$ is the Grassmann parity of $\Delta_\subb$).
The non-written terms in (\ref{S2}) encode analogously the
graded commutator algebra of the gauge symmetries, and of the
gauge symmetries with the $\Delta_\suba$.
Higher terms $S_k$ ($k>2$) in the expansion of $S$ reflect
consistency relations following from
the graded commutator algebra.
The solution of the extended master equation encodes thus
the complete algebraic structure of the gauge and rigid symmetries
under study.
In particular, the piece in $(S,S)=0$ which is linear in $\xi^*$
yields
\beq  
f_{[\suba\subb}^{\underline e}\, f_{\subc]\underline e}^{\underline d}=0
\label{jac}\eeq
where $[\dots]$ indicates graded antisymmetrization.
(\ref{jac}) is the Jacobi identity 
for the structure constants of a graded Lie algebra
and reflects again that the commutator algebra of the $\Delta_\suba$
constitutes a subalgebra of (\ref{S17}) in the soft sense.
Of course, in general this commutator algebra is not a 
true graded Lie algebra,
but still a graded Lie algebra in the soft sense.

\mysection{Extended BRST and Koszul-Tate differential}\label{KT}

The extended antifield formalism outlined in the previous
section implies the existence of
a nilpotent antiderivation which generalizes the standard
BRST differential so as to incorporate rigid symmetries.
We shall call this antiderivation the extended BRST differential
and denote it by $s$. It is defined
in the space of local functionals of the form (\ref{space})
via the extended antibracket through
\beq
sX=(S,X).
\label{s}\eeq
With this definition, $s$ squares to zero (= is ``nilpotent''),
\beq
s(XY)=(sX)Y+(-)^{\ep_X}X(sY),\quad s^2=0.
\label{lefts}\eeq
Furthermore, $s$ is a real differential if $S$ is a real functional.
As $s$ is Grassmann odd, this means, due to (\ref{cc}),
\beq
\overline{(sX)}=(-)^{\ep_X}s\bar X.
\eeq
It is useful to expand $s$ in the antifield number.
The structure of $S$ implies that the expansion of $s$ starts with a
piece $\delta$ that has antifield number $-1$
(i.e., $\delta$ lowers the antifield number by one unit),
\beq
s=\delta+\gamma+\sum_{i\geq 1} s_i\ ,\quad
\agh\delta=-1,\ \agh\gamma=0,\ \agh s_i=i .
\label{sdec}\eeq
The nilpotency of $s$ implies anticommutation relations between the
pieces in this decomposition,
\beq
\delta^2=0,\quad \mbox{{\bf [}}\delta,\gamma\mbox{{\bf ]}}=0,\quad
\gamma^2+\mbox{{\bf [}}\delta,s_1\mbox{{\bf ]}}=0,\quad \dots\quad .
\label{2}\eeq

$\delta$ is the extension of the field theoretical 
Koszul-Tate differential \cite{KT,proc,henteit}. It
acts nontrivially only on the antifields, and coincides on 
$\Phi^*_A$ with the standard Koszul-Tate differential,
while $\delta\xi^*_\suba$ is an integrated local functional
associated with the corresponding rigid symmetry,
\bea
& &\delta\Phi^A=\delta\xi^\suba=0\ ,\quad
\delta\phi^*_i=\frac{\7\6^R L}{\7\6\phi^i}\nonumber\\
& &\delta C^*_\alpha=R^i_\alpha{}^\dagger\phi^*_i\ ,
\quad \delta\xi^*_\suba =
(-)^{\ep_\suba}\int d^nx \, (\Delta_\suba\phi^i)\phi^*_i
\label{delta2}\eea
where $R^i_\alpha{}^\dagger$ is the
operator adjoint to $R^i_\alpha$ (its precise definition,
which includes a sign depending on the Grassmann parity, 
follows from the formulae).

$\delta$ is a nilpotent antiderivation by (\ref{2}). 
It therefore establishes the cohomological groups $H_k(\delta)$
at antifield number $k$
in the space of local functionals (\ref{space}).
By construction, $\delta$ is acyclic at all positive antifield
numbers ($H_k(\delta)\simeq 0$ $\forall k>0$)
when $S$ encodes {\em all} the gauge and rigid symmetries
(of first and higher order) \cite{bhw}.
In contrast, when only a subset of the rigid symmetries
is included, $H_k(\delta)$ corresponds at positive antifield number $k$
to the remaining rigid symmetries of order $k$ and is represented
by functionals that would be of the form $M^\resta(\xi)\delta\xi^*_\resta$
if all the rigid symmetries had been included.
Hence, $H_1(\delta)$ is represented by functionals
\beq
M^{\resta}(\xi)\int d^nx \, (\Delta_\resta\phi^i)\phi^*_i\ .
\label{delta3}\eeq

\mysection{Deformation theory}\label{DEF}

The extended antifield formalism allows one
to describe deformations of a given model and some of its
symmetries as deformations of the solution of the extended master
equation along the lines of \cite{bh}.
However, as anticipated in the introduction, a
deformation does not necessarily preserve the property that the selected
subset of symmetries is a closed one. Therefore, the deformation
itself may make it necessary to enlarge the subset of symmetries
one has started with. In this section we describe how to cope with this
phenomenon within a systematic approach to the deformation problem.

The starting point is a solution $\UD0S$ of the extended master 
equation which encodes the original (undeformed) classical action, its 
gauge symmetries and a closed subset $\{{\UD0\Delta}_\suba\}$ 
of its rigid symmetries. The basic idea is to
seek a continuous deformation of this solution of the form
\beq
S=\UD 0S+g\UD 1S+g^2\UD 2S+\dots
\label{D2}\eeq
where $g$ is the deformation parameter.
This problem is analysed ``perturbatively''
by expanding $(S,S)=0$ in $g$,
\bea
& (\,\UD0S,\UD0S\,)=0 &
\label{D3}\\
& (\,\UD0S,\UD1S\,)=0 &
\label{D4}\\
& (\,\UD1S,\UD1S\,)+2\,(\,\UD0S,\UD2S\,)=0 &
\label{D5}\\
& \vdots &
\nonumber
\eea
(\ref{D3}) is satisfied by assumption. In order to
discuss the subsequent equations,
one may now be tempted to adopt the arguments valid
for deformations preserving only the gauge symmetries, 
as given in \cite{bh}. One would then conclude from (\ref{D4})
that $\UD1S$ must be invariant under the undeformed
extended BRST differential $\UD0s$, as the latter 
is generated by the antibracket with $\UD0S$, see (\ref{s}).
Furthermore one can assume without loss of generality that
$\UD1S$ is nontrivial in the cohomology of
$\UD0s$, because otherwise it can be removed through local
field redefinitions and/or redefinitions of the
gauge and rigid symmetry transformations by adding trivial
symmetries.
This follows from standard arguments which parallel those
for deformations of gauge symmetries (see e.g.\ \cite{henmoscow})
and are not repeated here. In this way one would
conclude that $\UD1S$ represents a
nontrivial cohomology class of $H^0(\UD0s)$, the cohomology of
$\UD0s$ at ghost number 0 in the space of local
functionals (\ref{space}). However, this kind of reasoning overlooks 
that $\UD0s$ encodes only a subset of the rigid symmetries and
may thus be extended, if necessary.

In order to discuss this possibility, 
we analyse (\ref{D4}) and the subsequent equations more
carefully by expanding them in the antifield number. To this end 
we denote the decomposition of $\UD{n}S$ by
\beq
\UD{n}S=\sum_{k\geq 0}\UD{n}{S_k}\ ,\quad \agh\UD{n}{S_k}\,=k.
\label{D6}\eeq
The interpretation of the various terms in this expansion follows
from the general discussion in section \ref{AF}:
$\UD{n}{S_0}$ is the deformation of the original classical
action at order $n$ in $g$, $\UD{n}{S_1}$ encodes the
$n$th order deformations of the symmetry transformations under study,
$\UD{n}{S_2}$ yields the $n$th order deformation of the graded
commutator algebra of these symmetries etc..
Using the expansion of $\UD0s$ in the antifield number
as in (\ref{sdec}), Eq.\ (\ref{D4}) decomposes into
\bea
& \UD0\gamma\, \UD1{S_0}+\UD0\delta\, \UD1{S_1}\,=0 &
\label{D7}\\
& \UD0{s_1}\, \UD1{S_0}+\UD0\gamma\, \UD1{S_1}
+\UD0\delta\, \UD1{S_2}\,=0 &
\label{D8}\\
& \vdots &
\nonumber
\eea
(\ref{D7}) requires $\UD1{S_0}$
to be {\em invariant on-shell} under the undeformed gauge and rigid
symmetries under study, where ``on-shell'' refers to the undeformed
equations of motion. This is so because the undeformed symmetries
under study and the original equations of motion
are encoded in $\UD0\gamma$ and $\UD0\delta$ respectively.
Let us assume we have found a solution to (\ref{D7}).
The possible need for an enhancement of the subset of
rigid symmetries under study arises for the first time in the
next step, i.e., when seeking a solution of (\ref{D8}).
To see this we act with $\UD0\gamma$ on (\ref{D7}).
Using the anticommutation relations (\ref{2}) for $\UD0s$, we infer
that the functional $W_1$ defined by
\beq
W_1=\,\UD0{s_1}\, \UD1{S_0}+\UD0\gamma\, \UD1{S_1}
\label{W1}\eeq
is $\UD0\delta$-closed,
\beq
\UD0\delta W_1=0.
\label{D9}\eeq
Now, (\ref{D8}) requires that $W_1$ be $\UD0\delta$-exact.
(\ref{D9}) is thus a necessary condition
for the existence of a solution to (\ref{D8}).
However, it is not sufficient in general
when $\UD0s$ encodes only a subset of the
rigid symmetries, see section \ref{KT}.

The question 
at this stage is therefore: can it happen that $W_1$ contains
a rigid symmetry of the original action which is not contained
in the closed subset of symmetries one has started with? The
answer to this question is affirmative, as we shall
illustrate explicitly in the next sections.
Hence, as $W_1$ has antifield number 1, it may contain contributions 
of the form (\ref{delta3}). Furthermore $W_1$ has ghost number 1.
Its general form is thus
\beq
W_1=\frac 12\, (-)^{\ep_\suba+\ep_{\restc}}
\UD1{f_{\suba\subb}^{\restc}} \xi^\subb\xi^\suba
\int d^nx\, (\, {\UD0\Delta}_{\restc}\phi^i\, )\, \phi^*_i
-\UD0\delta (\dots).
\label{D10}\eeq
Recall that ${\UD0\Delta}_{\restc}$ denotes a rigid symmetry
of $\UD0{S_0}$ that is not contained in $\{{\UD0\Delta}_\suba\}$. 
If such symmetries occur in $W_1$,
i.e., if there are nonvanishing coefficients 
$\UD1{f_{\suba\subb}^{\restc}}$,
the subset of rigid symmetries under study needs to be enlarged
by including these symmetries in order to sove (\ref{D8}). 
Of course, this requires first of all to
construct a new solution $\UD0S$ of the extended master equation
which incorporates the additional symmetries too, and then
to reexamine Eqs.\ (\ref{D7}) and (\ref{D8})
as $\UD0s$ gets extended.

Let us assume now that Eqs.\ (\ref{D7}) and (\ref{D8}) have been solved.
Then there are no further obstructions to a solution of Eq.\ (\ref{D4})
if higher order symmetries are absent, i.e.,
all the equations subsequent to (\ref{D8}) can be solved without
further ado because then $\UD0\delta$ is acyclic at all antifield
numbers exceeding 1. In contrast, if there are
higher order symmetries, it cannot be excluded in principle
that some of them show up at a
certain stage and must be included too.

Once one has solved (\ref{D4}), one has to analyse
(\ref{D5}) and the subsequent equations. Now, one has
$(\,\UD0S,(\,\UD1S,\UD1S\,))=0$ as a consequence of (\ref{D4}),
thanks to the Jacobi identity for the extended antibracket.
$(\,\UD1S,\UD1S\,)$ has ghost number 1 and is thus a cocycle in
$H^1(\UD0s)$.
This is a necessary condition for the existence of a solution to
(\ref{D5}) but, in general, it is not sufficient because
(\ref{D5}) requires that
$(\,\UD1S,\UD1S\,)$ be $\UD0s$-exact. Therefore
(\ref{D5}) may obstruct deformations
through $H^1(\UD0s)$. Note however that some of the
cohomology classes in $H^1(\UD0s)$ will originate
from rigid symmetries that have not been included so far. These
classes are represented by $\UD0s$-invariant extensions of
functionals of the form (\ref{delta3}) and their analogues
for higher order rigid symmetries (if any). Such classes 
can be removed by further extending the subset of rigid symmetries.
We shall therefore refer to them as ``spurious anomalies'', and
call the other classes ``true anomalies''.%
\footnote{The term ``anomaly'' is (ab)used here because these
obstructions parallel those to the Slavnov-Taylor
identity through gauge anomalies in quantum field theory.
Indeed, the Slavnov-Taylor identity can be cast in the form 
of the master equation \cite{zj,Lee} and the gauge anomalies 
represent BRST cohomology classes at ghost number 1 \cite{brs}.}
These two kinds of anomalies show up at different
antifield numbers%
\footnote{I have not found an example where 
$(\,\UD1S,\UD1S\,)$ contains spurious anomalies. On the other
hand, I have neither found a general argument which excludes the 
occurrence of spurious anomalies. Hence, the question whether
or not such anomalies can really occur in $(\,\UD1S,\UD1S\,)$
is actually still open.}. 
Using the expansion
\beq
(\,\UD1S,\UD1S\,)=-2\sum_{k\geq 0} A_k\quad,\quad\agh A_k=k,
\eeq
(\ref{D5}) decomposes into
\bea
& A_0=\UD0\gamma\,\UD2{S_0}\,+\UD0\delta\,\UD2{S_1}&
\label{D99}\\
& A_1=\UD0{s_1}\,\UD2{S_0}\,+
\UD0\gamma\,\UD2{S_1}\,+\UD0\delta\,\UD2{S_0}&
\label{D100}\\
& \vdots &
\nonumber
\eea
True anomalies can show up only in $A_0$ through contributions that are
weakly (= on-shell) $\UD0\gamma$-closed but not weakly
$\UD0\gamma$-exact. They can thus obstruct (\ref{D99}). In contrast,
spurious anomalies would show up in the $A_k$ with $k>0$, and thus in
(\ref{D100}) and the equations subsequent to it.
Thereby spurious anomalies stemming from rigid symmetries
of order $k$ would show up in $A_k$. In particular, when higher
order rigid symmetries are absent, actually only $A_1$ can
give rise to spurious anomalies through terms of the form
(\ref{delta3}) with ghost number 1.
Analogously one analyses the equations
subsequent to (\ref{D5}) and infers that they can obstruct
the deformation in the same way through $H^1(\UD0s)$.

To summarize, the extended antifield formalism permits a systematic
analysis of deformations preserving certain
rigid symmetries in addition to the gauge symmetries in a manner
which is quite similar to the deformation theory \cite{bh} based on the
standard antifield formalism. The main difference is
that the deformation
itself may force one to enlarge the subset of
rigid symmetries one has started with.
It should be clear from the above discussion that
in general one cannot predict from the outset which symmetries need 
to be included in addition to those one has started with
because that may depend on the solution to (\ref{D7}).

A deformation which requires the enlargement
of an originally closed subset of symmetries
results in a deformed symmetry algebra. For instance,
(\ref{D10}) would yield
\beq
\UD1{S_2}\, =
\frac 12\, (-)^{\ep_\subb+1}\xi^\subb \xi^\suba 
\UD1{f_{\suba\subb}^{\restc}}\xi^*_{\restc}
+\dots\quad .
\label{D11}\eeq
This shows that the graded commutator algebra of the
$\Delta_\suba$ (i.e., of the deformed transformations) would not close
anymore in the soft sense but involve the $\Delta_{\restc}$.
The $\UD1{f_{\suba\subb}^{\restc}}$ are
the corresponding structure constants of the deformed graded Lie 
algebra to first order.

\mysection{Central charge of the N=2 hypermultiplet}\label{HM}

As an illustration, we shall now treat an
N=2 supersymmetric model for a Fayet-Sohnius hypermultiplet
\cite{Fa,So} in flat four dimensional spacetime.
The multiplet contains two complex Lorentz-scalar fields $\varphi^i$
($i=1,2$) and two complex Weyl-spinor fields $\chi^\alpha$, 
$\psi^\alpha$ ($\alpha=1,2$)%
\footnote{We use conventions
with a Minkowski metric $\eta_{\mu\nu}=diag(+,-,-,-)$
as in \cite{sugra}
which differ only through signs from those in \cite{wb}.}.
As basis of the classical fields we use these fields and their 
complex conjugates (equivalently we could have chosen for instance 
the real and imaginary parts of the fields),
\[ 
\{\phi\}=\{\varphi^i\, , \, \5\varphi_i\, , \, \chi^\alpha\, , \, 
\5\chi^\da\, , \, \psi^\alpha\, , \, \5\psi^\da\}
\]
where $\5\varphi_i$, $\5\chi$ and
$\5\psi$ are complex conjugate to $\varphi^i$, $\chi$ and
$\psi$ respectively,
\[
\5\varphi_i=\overline{\varphi^i},\quad
\5\chi^\da=\overline{\chi^\alpha},\quad
\5\psi^\da=\overline{\psi^\alpha}.
\]
The position of the index of $\5\varphi_i$ indicates 
that it transforms contragrediently to $\varphi^i$ 
under the SU(2)-automorphism group of N=2 supersymmetry ($i$
refers to the fundamental representation of this SU(2)).
Undotted and dotted spinor indices distinguish the (1/2,0) and
(0,1/2) representations of the Lorentz group (resp.,\ of its
covering group SL(2,C)).

Our starting point is the action
\bea
\UD0{S_0} &=& \int d^4x \left[\6_\mu \varphi^i \6^\mu \5\varphi_i
-\frac \Ii 2\,(\chi\dsl\5\chi+\5\chi\dsl\chi
+\psi\dsl\5\psi+\5\psi\dsl\psi)\right]
\label{N1}\eea
where
\[\dsl_{\alpha\da}=\sigma^\mu_{\alpha\da}\6_\mu\ .\]
The action $\UD0{S_0}$ is among others invariant under
rigid N=2 supersymmetry
transformations ${\UD0\Delta}_\alpha{}^i$,
${\UD0\Delta}_{\da i}$ given by
\beq
\ba{c|cccccc}
\phi & \varphi^j & \5\varphi_j & \chi^\beta & 
\5\chi^{\dot\beta} & \psi^\beta & \5\psi^{\dot\beta}
\\
\hline\rule{0em}{3ex}
{\UD0\Delta}_\alpha{}^i\phi
& \ep^{ij}\chi_\alpha & \delta^i_j\psi_\alpha & 
0 & -\Ii\!\dsl_\alpha^{\dot\beta}\5\varphi^i & 0 & 
-\Ii\!\dsl_\alpha^{\dot\beta}\varphi^i
\\
\hline\rule{0em}{3ex}
{\UD0\Delta}_{\da i}\phi
& \delta_i^j\5\psi_\da & -\ep_{ij}\5\chi_\da & 
\Ii\!\dsl_\da^\beta\varphi_i & 0 & -\Ii\!\dsl_\da^\beta\5\varphi_i & 0
\ea\label{N2}\eeq
where indices $i$
are raised and lowered with the rules
\[
X^i=\ep^{ij}X_j\ ,\ X_i=\ep_{ij}X^j\ ,\
\ep^{ij}=-\ep^{ji}\ ,\ \ep_{ij}=-\ep_{ji}\ ,\ 
\ep^{12}=\ep_{21}=1.\]
We consider the following subset of rigid symmetries,
containing the supersymmetry transformations and the
spacetime translations,
\beq
\{{\UD0\Delta}_\suba\}\equiv\{{\UD0\Delta}_\alpha{}^i\, ,\,
{\UD0\Delta}_{\da i}\, ,\, \6_\mu\}.
\label{N3}\eeq
The graded commutator algebra of these symmetries reads
\beq
\mbox{{\bf [}}\, 
{\UD0\Delta}_\alpha{}^i\, ,\, {\UD0\Delta}_{\da j}\, 
\mbox{{\bf ]}}
\approx -\Ii\,\delta_j^i\dsl_{\alpha\da}\quad ,\quad
\mbox{{\bf [}}\, 
{\UD0\Delta}_\suba\, ,\, {\UD0\Delta}_\subb\, 
\mbox{{\bf ]}}
\approx 0\quad\mbox{otherwise}
\label{N4}\eeq
where $\approx$ denotes equality up to on-shell trivial
symmetries. (\ref{N4}) is indeed the N=2 supersymmetry
algebra without central charge (on-shell).
The action has no gauge symmetries.
Therefore it has no higher order symmetries either \cite{bbh1}.
This implies the existence of a solution
to the extended master equation which encodes only the
symmetries (\ref{N3}) and their graded commutator algebra.
This solution, which was computed first in \cite{bau}, reads
\bea
\UD0S &=& \UD0{S_0}+\UD0{S_1}+\UD0{S_2}
\nonumber\\
\UD0{S_1} &=& -\int d^4x\sum_\phi 
(\xi^\alpha_i{\UD0\Delta}_\alpha{}^i\phi
+\5\xi^{\da i}{\UD0\Delta}_{\da i}\phi
+\xi^\mu\6_\mu \phi)\,\phi^*
\nonumber\\
\UD0{S_2} &=& -\Ii\,\xi_i\sigma^\mu\5\xi^i\,\xi^*_\mu
+\int d^4x
\left[
\5\chi^*\5\xi^i\,\xi_i\chi^*
+\5\psi^*\5\xi^i\,\xi_i\psi^*
\right.
\nonumber\\
& &
\left.
+\frac 12\,\xi_i\xi^i\, \5\chi^*\5\psi^*
+\frac 12\,\5\xi^i\5\xi_i\, \psi^*\chi^*
\right]
\label{N5}\eea
where 
\[ 
\{\phi^*\}=\{\varphi^*_i\, , \, \5\varphi^{i*}\, , \, 
\chi^*_\alpha\, , \, \5\chi^*_\da\, , \,
\psi^*_\alpha\, , \, \5\psi^*_\da\}.
\]
The supersymmetry ghosts $\xi^\alpha_i$ and $\5\xi^{\da i}$
are Grassmann even and the translation ghosts $\xi^\mu$ are
Grassmann odd. The ghosts and antifields have the reality properties
\beann 
& \5\xi^{\da i}=\overline{\xi^\alpha_i}\quad ,\quad
\xi^\mu=\overline{\xi^\mu}\quad ,\quad
\5\xi^*_{\da i}=-\overline{\xi^{i*}_\alpha}\quad ,\quad
\xi^*_\mu=-\overline{\xi^*_\mu} &
\\
& \5\varphi^{i*}=-\overline{\varphi^*_i}\quad ,\quad
\5\chi^*_\da=-\overline{\chi^*_\alpha}\quad ,\quad
\5\psi^*_\da=-\overline{\psi^*_\alpha}\quad . &
\eeann
The first term in $\UD0{S_2}$ contains the structure constants
of the supersymmetry algebra (\ref{N4}), while
the contributions which are quadratic in the antifields reflect 
that the symmetry algebra closes only on-shell.

We now study deformations of the above model along
the lines of the previous section.
A solution to (\ref{D7}) which introduces mass terms
for the fermions is easily found. Namely,
\beq
\UD1{S_0}\, =  \int d^4x \left[ m_1\chi\psi+\5m_1\5\chi\5\psi
+\frac 12\, (m_2\chi\chi+\5m_2\5\chi\5\chi
+m_3\psi\psi+\5m_3\5\psi\5\psi)\right]
\label{N6}\eeq
is supersymmetric on-shell and translation invariant 
for any choice of complex mass parameters
$m_1$, $m_2$ and $m_3$ and therefore yields a solution to
(\ref{D7}). The corresponding functional $\UD1{S_1}$ is
\bea
\UD1{S_1} =
\int d^4x \left[
m_1(\5\varphi_i\5\chi^*\5\xi^i-\varphi_i\5\psi^*\5\xi^i)
-\5m_1(\5\varphi^i\xi_i\psi^*+\varphi^i\xi_i\chi^*)
\right.
\nonumber\\
\left.
-m_2\varphi_i\5\chi^*\5\xi^i-\5m_2\5\varphi^i\xi_i\chi^*
+m_3\5\varphi_i\5\psi^*\5\xi^i-\5m_3\varphi^i\xi_i\psi^*
\right].
\label{N7}\eea
Next we calculate the functional $W_1$ in (\ref{W1}).
The result is
\bea
\UD0{s_1}\, \UD1{S_0} +\UD0\gamma\, \UD1{S_1}\, =
\frac 12\,\5\xi^i\5\xi_i \int d^4x \left[
(-m_1\varphi^j+m_3\5\varphi^j)\varphi_j^*
+\5\varphi^{j*}(m_1\5\varphi_j-m_2\varphi_j)
\right.
\nonumber\\
-(m_1\chi+m_3\psi)\chi^*+\5\chi^*(m_1\5\chi-m_2\5\psi)
\nonumber\\
\left.
+(m_1\psi+m_2\chi)\psi^*+\5\psi^*(-m_1\5\psi+m_3\5\chi)\right]
+\mbox{c.c.}
\label{N8}\eea
where c.c. denotes complex conjugation. (\ref{N8}) has the
form of the first term in (\ref{D10}), i.e., it brings in
an additional symmetry. This symmetry
is part of a rigid SU(2)-invariance of the
action (\ref{N1}). Indeed, as the functional (\ref{N8})
is $\UD0\delta$-invariant for any choice of
$m_1$, $m_2$ and $m_3$, the parts in (\ref{N8})
involving $m_1$, $m_2$ and $m_3$ respectively
correspond to independent symmetries of the action (\ref{N1}). 
These symmetries form an SU(2) under which
$(\varphi^1,\5\varphi^1)$, $(\varphi^2,\5\varphi^2)$,
$(\chi,\psi)$ and $(\5\psi,\5\chi)$ transform as doublets
(i.e., in the fundamental representation) and which
commutes with the supersymmetry transformations (\ref{N2}).
However, in contrast to the undeformed action, the
first order deformation (\ref{N6})
is not invariant under the full SU(2) but it is
still invariant under 
a U(1) subgroup thereof generated by the
transformations in (\ref{N8}). Hence, the deformation breaks
the SU(2) but preserves this U(1) subgroup%
\footnote{An analogous phenomenon was
observed in \cite{WHP} within the construction of supergravity 
couplings for hypermultiplets.}.

We thus have to enlarge the subset of symmetries
(\ref{N3}) by this U(1). It turns out that this
suffices in order to construct a deformed
solution of the extended master equation.
We shall not further discuss the computation and 
spell out the solution only for the case
$m_2=m_3=0$. Using $m=g m_1$ ($g$ being the deformation
parameter in the notation of the previous section),
the deformed solution reads then
\bea
S &=& S_0 + S_1 + S_2
\label{N9}\\
S_0 &=& \UD0{S_0} + 
\int d^4x\,(m\chi\psi+\5m\5\chi\5\psi-m\5m\varphi^i\5\varphi_i)
\label{N10}\\
S_1 &=& \int d^4x\, [-\sum_\phi 
(\xi^\alpha_i{\UD0\Delta}_\alpha{}^i\phi
+\5\xi^{\da i}{\UD0\Delta}_{\da i}\phi
+\xi^\mu\6_\mu \phi)\,\phi^*
\nonumber\\
& & +\Ii\, \xi_{U(1)} 
(\varphi^i\varphi_i^*-\5\varphi_i\5\varphi^{i*}
+\chi\chi^*-\5\chi^*\5\chi-\psi\psi^*+\5\psi^*\5\psi)
\nonumber\\
& &
+m(\5\varphi_i\5\chi^*\5\xi^i-\varphi_i\5\psi^*\5\xi^i)
-\5m(\5\varphi^i\xi_i\psi^*+\varphi^i\xi_i\chi^*)]
\label{N11}\\
S_2 &=& \UD0{S_2} 
+\frac {\Ii}2\, (m\,\5\xi^i\5\xi_i+\5m\,\xi_i\xi^i)\, \xi^*_{U(1)}
\label{N12}
\eea
with $\UD0{S_2}$ as in (\ref{N5}).
Here $\xi_{U(1)}$ and $\xi^*_{U(1)}$ are the global ghost and antifield
of the rigid U(1) symmetry obtained from (\ref{N8}) in the case
$m_2=m_3=0$ ($\xi_{U(1)}$ is real and Grassmann odd, 
$\xi^*_{U(1)}$ is purely
imaginary and Grassmann even).

(\ref{N10}) is the deformed classical action. Apart from the
original action (\ref{N1}) and its first order deformation
(\ref{N6}) (in the case $m_2=m_3=0$), it contains also a mass term
for the Lorentz-scalar fields which arises at second order
in the deformation parameter.

(\ref{N11}) contains the deformed supersymmetry transformations,
the rigid U(1) transformations,
and the spacetime translations. 
The deformed supersymmetry and the U(1) transformations are
\beq
\ba{c|cccccc}
\phi & \varphi^j & \5\varphi_j & \chi^\beta & 
\5\chi^{\dot\beta} & \psi^\beta & \5\psi^{\dot\beta}
\\
\hline\rule{0em}{3ex}
\Delta_\alpha{}^i\phi
& \ep^{ij}\chi_\alpha & \delta^i_j\psi_\alpha & 
\5m\delta_\alpha^\beta\varphi^i & 
-\Ii\!\dsl_\alpha^{\dot\beta}\5\varphi^i & 
\5m \delta_\alpha^\beta\5\varphi^i & 
-\Ii\!\dsl_\alpha^{\dot\beta}\varphi^i
\\
\hline\rule{0em}{3ex}
\Delta_{\da i}\phi
& \delta_i^j\5\psi_\da & -\ep_{ij}\5\chi_\da & 
\Ii\!\dsl_\da^\beta\varphi_i & 
-m\delta_\da^{\dot \beta}\5\varphi_i & 
-\Ii\!\dsl_\da^\beta\5\varphi_i & 
m\delta_\da^{\dot \beta}\varphi_i
\\
\hline\rule{0em}{3ex}
\Delta_{U(1)}\phi & -\Ii\varphi^j & \Ii\5\varphi_j & -\Ii\chi^\beta & 
\Ii\5\chi^{\dot\beta} & \Ii\psi^\beta & -\Ii\5\psi^{\dot\beta}
\ea\label{N13}\eeq

(\ref{N12}) encodes the graded commutator algebra of the
deformed symmetry transformations. The N=2 supersymmetry algebra
has become extended by the U(1) through the deformation. 
The nonvanishing graded commutators
are
\bea
\mbox{{\bf [}}\, 
\Delta_\alpha{}^i\, ,\, \Delta_{\da j}\, 
\mbox{{\bf ]}}
& \approx & -\Ii\,\delta_j^i\dsl_{\alpha\da} 
\nonumber\\
\mbox{{\bf [}}\, 
\Delta_\alpha{}^i\, ,\, \Delta_\beta{}^j\, \mbox{{\bf ]}}
& \approx & \Ii\, \5m\, \ep^{ij}\ep_{\alpha\beta}\Delta_{U(1)} 
\nonumber\\
\mbox{{\bf [}}\, 
\Delta_{\da i}\, ,\, \Delta_{\dot \beta j}\, 
\mbox{{\bf ]}}
& \approx & -\Ii\, m\, \ep_{ij}\ep_{\da\dot \beta}\Delta_{U(1)} 
\label{N14}\eea
where $\approx$ now denotes equality up to transformations which
are trivial on-shell in the deformed model (i.e., these
transformations involve the deformed equations of motion).

{\sc Remark.} The above results hold analogously in a formulation
of the hypermultiplet with the standard auxiliary fields used
already in \cite{Fa,So}. In that approach one sometimes introduces an
``off-shell central charge'' in order to close the commutator
algebra of the supersymmetries, the central charge and the
spacetime translations off-shell. However, in the massless model
that central charge is trivial on-shell and thus not to be
accompanied by global ghosts. In contrast, the
massive (deformed) model involves again a ``true'' central 
charge that does not vanish on-shell.

\mysection{Curci-Ferrari model}\label{CF}

A particular case of a rigid symmetry is the BRST symmetry
of a gauge fixed action constructed in the standard
way from a solution to the usual master equation 
\cite{bv,henteit,report}. Deformations of a gauge fixed action
may be obtained in two ways: (i) one
constructs first consistent deformations of the
underlying gauge theory along the
lines of \cite{bh} and fixes the gauge afterwards, or
(ii) one investigates directly deformations of the 
gauge fixed model and its BRST symmetry.

These two approaches are not equivalent in general. In particular,
the first approach leads by construction to an on-shell nilpotent BRST 
symmetry of the standard type,
whereas the second one may destroy the nilpotency
property and is not physically acceptable in general.
This is related to the different properties of the BRST cohomology
before and after gauge fixing \cite{henfix} (cf.\ also the remark
at the end of this section),
and is now to be discussed for the Curci-Ferrari model
\cite{CF,O,kostas} in the framework of
the extended antifield formalism. The loss of nilpotency 
emerges in this approach as a deformation of the
BRST algebra along the lines of section \ref{DEF}: in this particular 
case, the deformed action has even the {\em same} BRST symmetry as 
the original one, but in the deformed model that symmetry does not
square weakly to zero anymore (as the equations of motion
change). Rather, it squares into a different nontrivial 
rigid symmetry.

We consider four dimensional nonabelian 
Yang--Mills theory with the following
gauge fixed action%
\footnote{The gauge fixed action (\ref{C1}) arises in the standard 
manner from a ``minimal'' solution $S_{min}$ of the usual master equation
as follows. First one adds to $S_{min}$ the ``nonminimal'' term
$\int d^4x\, \mbox{Tr}(-H B^*)$ where the $H^i$ are
Nakanishi-Lautrup auxiliary fields. Then one shifts the antifields by
$\Phi^*_A\rightarrow\Phi^*_A+\delta^L\Psi/\delta\Phi^A$ where
$\Psi$ is the ``gauge fixing fermion'' $\Psi[\Phi]=
\int d^4x\, \mbox{Tr} [(\alpha/2)(BH+eB^2C)-B\6_\mu A^\mu]$.
Finally one eliminates the $H^i$ by their algebraic equations
of motion.},
\beq
\UD0{S_0}\, = \int d^4x\,  \mbox{Tr} \left[
\frac 14\, F_{\mu\nu}F^{\mu\nu}+\frac 1{2\alpha}\, (\6_\mu A^\mu)^2
-\frac 12\, B(\6_\mu D^\mu+D_\mu\6^\mu)C
+\frac {\alpha e^2}{4}\, B^2C^2
\right]
\label{C1}\eeq
where $\alpha$ is the gauge fixing parameter, $e$ is the
gauge coupling constant, and
$A_\mu=A_\mu^i t_i$, $F_{\mu\nu}=F_{\mu\nu}^i t_i$, 
$C=C^i t_i$ and $B=B^i t_i$ are the Lie algebra valued
gauge fields, field strengths, ghost fields and 
antighost fields respectively ($\{t_i\}$ denotes an
appropriate matrix representation of the Lie algebra
of the gauge group normalized such that 
$\mbox{Tr}(t_it_j)=-\delta_{ij}$), and $D_\mu C$ is defined by
\beq
D_\mu C = \6_\mu C + e\, (A_\mu C-C A_\mu).
\label{C2}\eeq 
The action (\ref{C1}) is invariant under the rigid 
BRST transformations
\beq
\Delta_{\mbox{brs}} A_\mu = D_\mu C,\quad 
\Delta_{\mbox{brs}} C = -e\, C^2,\quad 
\Delta_{\mbox{brs}} B = \frac 1\alpha\, \6_\mu A^\mu 
- \frac e2\, (BC+CB).
\label{C3}\eeq
These transformations are nilpotent on-shell,
\beq
(\Delta_{\mbox{brs}})^2
=\frac 12\, \mbox{{\bf [}}\, \Delta_{\mbox{brs}},
\Delta_{\mbox{brs}}\, \mbox{{\bf ]}}
\approx 0.
\label{C4}\eeq
More precisely, $\Delta_{\mbox{brs}}$ is strictly
nilpotent on $A_\mu^i$ and $C^i$, but squares into an on-shell
trivial symmetry on $B^i$,
\beq
(\Delta_{\mbox{brs}})^2 A_\mu^i=
(\Delta_{\mbox{brs}})^2C^i=0,\quad
(\Delta_{\mbox{brs}})^2B^i= \frac 1\alpha\,
\delta^{ij}\, \frac{\delta^L\UD0{S_0}}{\delta B^j}\ .
\label{C4a}\eeq
We shall now apply the extended antifield formalism to the 
gauge fixed action (\ref{C1}) and the BRST symmetry (\ref{C3}). 
In this approach (\ref{C1}) plays the role of the classical 
action, i.e.\ the ghost and antighost fields $C$ and $B$ are 
viewed as Grassmann odd ``classical'' fields ($C$ is real,
$B$ purely imaginary),
\[
\{\phi\}=\{A_\mu^i\, ,\, B^i\, ,\, C^i\}
\]
Accordingly, we assign antifield number 1 to
$A^{\mu *}_i$, $B^*_i$ and $C^*_i$.
Furthermore we introduce a Grassmann even global ghost 
$\xi_{\mbox{brs}}$ for the
BRST symmetry (\ref{C3}). 

That is, in this case we consider a
subset of rigid symmetries containing only one element, namely 
$\Delta_{\mbox{brs}}$:
\beq
\{{\UD0\Delta}_\suba\}=\{ \Delta_{\mbox{brs}} \}.
\eeq
The corresponding graded commutator algebra (\ref{S17})
is just (\ref{C4a}). As the
gauge fixed action (\ref{C1}) has no gauge symmetry and thus
no higher order rigid symmetry either,
a corresponding solution of the extended master equation exists.
This solution coincides of course with the gauge fixed solution of
the master equation obtained in the standard antifield formalism, except
that now the global ghost $\xi_{\mbox{brs}}$ appears,
\bea
\UD0S &=& \UD0{S_0}+\UD0{S_1}+\UD0{S_2}
\nonumber\\
\UD0{S_1} &=& \xi_{\mbox{brs}}\int d^4x\,\mbox{Tr} \left[
A^{\mu*}D_\mu C +e\, C^2 C^*-\left\{ \frac 1\alpha\, \6_\mu A^\mu 
- \frac e2\, (BC+CB)\right\} B^*\right]
\nonumber\\
\UD0{S_2} &=&\frac 1{2\alpha}\, \xi_{\mbox{brs}}^2
\int d^4x\, \mbox{Tr}(B^* B^*)
\label{C5}
\eea
where we have used $A^{\mu*}=-\delta^{ij}A_i^{\mu*}t_j$ etc..
The presence of the term quadratic in $B^*$ reflects 
that the algebra closes on $B^i$ only on-shell, see (\ref{C4a}).
The ``extended'' BRST differential $\UD0s$, constructed from
$\UD0S$ as in Eq.\ (\ref{s}), coincides with the
usual gauge fixed BRST operator for the action (\ref{C1}), except
that now $\xi_{\mbox{brs}}$ occurs. It is strictly nilpotent, in contrast to 
$\Delta_{\mbox{brs}}$, and acts on the fields by
\bea
\UD0s A_\mu &=& \xi_{\mbox{brs}}\, D_\mu C,\quad 
\UD0s C =-\xi_{\mbox{brs}}\, e\, C^2
\nonumber\\
\UD0s B &=& \xi_{\mbox{brs}} \left[\frac 1\alpha\, \6_\mu A^\mu 
- \frac e2\, (BC+CB)\right]
-\frac 1{\alpha}\,\xi_{\mbox{brs}}^2 B^*. 
\label{C6}\eea

We shall now discuss the deformation of the action (\ref{C1})
through the Curci-Ferrari mass term
\beq
\UD1{S_0}\, =\int d^4x\, \mbox{Tr}
\left[ \frac 12\, A_\mu A^\mu+\alpha BC \right].
\label{C7}\eeq
This term is off-shell invariant under the transformations (\ref{C3}) 
and thus yields a solution to Eq.\ (\ref{D7}) with
\beq
\UD1{S_1}\, = 0.
\label{C8}\eeq
The functional $W_1$ in Eq.\ (\ref{W1}) reads in this case
\beq
\UD0{s_1}\UD1{S_0}\, =-\xi_{\mbox{brs}}^2\int d^4x\, \mbox{Tr} (B^*C).
\label{C8a}\eeq
This has the form of the first term in Eq.\ (\ref{D10}) and
contains an additional nontrivial rigid symmetry of the gauge fixed action
(\ref{C1}), namely
\beq
\Delta_{\mbox{add}}B=C,\quad 
\Delta_{\mbox{add}}A_\mu=\Delta_{\mbox{add}}C=0.
\label{C9}\eeq
Hence, in order to construct a deformed solution of the extended master
equation with $\UD1{S_0}$ as in (\ref{C7}), we {\em must} include 
this symmetry. It is straightforward to verify that
this yields the following deformed solution of the extended master
equation,
\bea
S&=&S_0+S_1+S_2
\nonumber\\
S_0&=&\UD0{S_0}+g\UD1{S_0}
\nonumber\\
S_1&=&-\int d^4x\sum_\phi\left(
\xi_{\mbox{brs}}\Delta_{\mbox{brs}} \phi+
\xi_{\mbox{add}}\Delta_{\mbox{add}} \phi\right)\phi^*
\nonumber\\
S_2&=&\UD0{S_2}+g\xi_{\mbox{brs}}^2\xi^*_{\mbox{add}}\ .
\label{C10}\eea
with $\UD0{S_2}$ as in (\ref{C5}).
The last term in (\ref{C10}) reflects that
the graded commutator algebra of $\Delta_{\mbox{brs}}$ 
and $\Delta_{\mbox{add}}$ reads in the deformed model
\beq
(\Delta_{\mbox{brs}})^2\approx g\Delta_{\mbox{add}}\quad,
\quad 
\mbox{{\bf [}}\, 
\Delta_{\mbox{brs}},\Delta_{\mbox{add}}\,
\mbox{{\bf ]}}=0
\label{C11}\eeq
where $\approx$ now denotes on-shell equality in the deformed model,
i.e.\ for the equations of motion following from
$\UD0{S_0}+g\UD1{S_0}$. Notice that $\Delta_{\mbox{brs}}$
is still a symmetry of the deformed model, without having been deformed.
Nevertheless it is not nilpotent anymore on-shell
because the equations of motion have changed.
\medskip

{\sc Remarks.} 

a) In order to avoid possible confusion, we stress
that the BRST cohomologies before and after gauge fixing are
always isomorphic (provided all the antifields are kept). 
What changes however when the gauge fixed action is
treated as a classical one, are the assignments of antifield
numbers and the corresponding concept of weak 
(= on-shell) equality (as the ghost fields count now among
the classical fields). As a consequence,
it is {\em not} true that each local functional with vanishing
antifield number which is on-shell 
$\Delta_{\mbox{brs}}$-invariant
can be extended to a cocycle of $\UD0s$
(this is just the phenomenon discussed in \cite{henfix}, but in the
language used here).
The Curci-Ferrari mass term (\ref{C7}) illustrates exactly
this phenomenon: it {\em cannot} be extended so as to be $\UD0s$-closed,
although it is $\Delta_{\mbox{brs}}$-invariant. As a consequence,
$\Delta_{\mbox{brs}}$ is not nilpotent anymore on-shell in the deformed
model.

b) The Curci-Ferrari model illustrates a general fact: 
a deformation of a gauge fixed action
which destroys the on-shell nilpotency of $\Delta_{\mbox{brs}}$
(or a deformation thereof) {\em cannot} reflect
a consistent deformation of the gauge symmetry in the sense
of \cite{bh} because such consistent deformations 
result by their very construction in
an on-shell nilpotent $\Delta_{\mbox{brs}}$ after gauge fixing.

c) Of course, (\ref{C10}) yields via (\ref{s})
a strictly nilpotent operator
which incorporates both $\Delta_{\mbox{brs}}$ and
$\Delta_{\mbox{add}}$. However, this nilpotent operator cannot
cure the unitarity problems of the Curci-Ferrari model discussed
in \cite{O,kostas} because $\Delta_{\mbox{add}}$
does not impose additional conditions that may select physical states.
Indeed, as it is just the square of $\Delta_{\mbox{brs}}$ 
(on-shell), a state that is annihilated by 
$\Delta_{\mbox{brs}}$ (resp.\ by its quantum version)
is automatically also annihilated by $\Delta_{\mbox{add}}$.
For the same reason, a state that is
$\Delta_{\mbox{add}}$-exact is also in the image of
$\Delta_{\mbox{brs}}$.

\mysection{Conclusion}\label{CON}

We have outlined how continuous
deformations of an action functional, its gauge symmetries and
a closed subset of its rigid
symmetries can be analysed systematically in
the extended antifield formalism.
The procedure is very similar to the study of 
continuous deformations
of actions and their gauge symmetries described in \cite{bh}. 
The main difference is that the deformation itself
may make it necessary to enlarge the particular subset of
rigid symmetries one has started with. This happens
when the commutator algebra of the deformed version of the 
originally considered subset of symmetries does not
close anymore in the soft sense (i.e.\ modulo gauge transformations
and on-shell trivial symmetries) and thus results in a
deformation of the symmetry algebra. 

It is however not always clear from the outset which
additional symmetries can occur in the deformed commutator
algebra. This subtlety can be mastered when one
proceeds as described in section \ref{DEF}, using an expansion
in the antifield number. In this approach
one first seeks functionals of the classical fields that are
``weakly'' (= on-shell) invariant under the symmetries under study. The
method then provides automatically the additional symmetries
which need to be included. This has been illustrated for
the hypermultiplet of four dimensional N=2 supersymmetry
where the central extension of the N=2 supersymmetry algebra
emerges via the deformation of a massless model to a massive one.
The central extension
turns out to be a surviving generator of an SU(2) symmetry of
the massless action broken by the deformation. 
In this case it depends on the mass parameters, i.e.\ on the deformation
itself, how the SU(2) is broken and which generator becomes the
central extension.

We have also illustrated, for the Curci-Ferrari model, how 
deformations of a
gauge fixed action and its BRST symmetry can be analysed 
within this approach.
The BRST symmetry is then treated in the same manner as
other rigid symmetries too, while the gauge fixed action
is treated as a classical one. 
However, such deformations do not correspond necessarily to
consistent deformations of the gauge symmetries in the sense
of \cite{bh}, and
are therefore not always physically acceptable.
In particular, it can happen that there are
deformations of a gauge fixed action which are BRST invariant
but nevertheless inconsistent because the BRST symmetry does not 
square to zero on-shell anymore in the deformed model. The 
Curci-Ferrari model illustrates exactly this phenomenon.
Hence, a necessary condition for a deformation of a gauge fixed
action to be a consistent one, is the on-shell nilpotency of the 
BRST symmetry of the deformed action.

Finally I remark that the procedure outlined in section \ref{DEF}
can be extended analogously to the case  that only a subset of
the gauge symmetries is included. However, from the physical
point of view this extension is mainly of academic
interest and was therefore not discussed here.
\medskip

\section*{Acknowledgements}

Discussions with Tobias Hurth and Kostas Skenderis about the
Curci-Ferrari model are gratefully acknowledged.
The author was supported by the Deutsche Forschungsgemeinschaft.

\end{document}